\documentclass[conference,usenatbib]{basi}
\usepackage[T1]{fontenc}
\usepackage[british]{babel}
\usepackage[varg]{txfonts}
\usepackage{rotating}
\usepackage{dcolumn}
\begin{document}
\title[Debnath et al.]{Characterization of few transient black hole candidates during their X-ray outbursts with TCAF Solution}
\author[Debnath et al.]%
       {Dipak Debnath$^1$\thanks{email: \texttt{dipak@csp.res.in}},
       S. Mondal$^1$, S. K. Chakrabarti$^{1,2}$, A. Jana$^1$, A. A. Molla$^1$, D. Chatterjee$^1$\\
       $^1$ Indian Centre for Spacce Physics, 43 Chalantika,Garia Station Road,Kolkata-700084,India\\
       $^2$ S. N. Bose National Center for Basic Sciences, JD-Block, Salt Lake, Kolkata, 700098, India}
\pubyear{2015}
\volume{12}
\pagerange{87-88}


\maketitle

\label{firstpage}

\begin{abstract}

The theoretical concept of Chakrabarti-Titarchuk two Component Advective Flow (TCAF) model was introduced around 
two decade ago in mid-90s. Recently after the inclusion of TCAF model into XSPEC as an additive table model, we 
find that it is quite capable to fit spectra from different phases of few transient black hole candidates (TBHCs) 
during their outbursts. This quite agrees with our theoretical understanding. Here, a brief summary of our recent studies 
of spectral and temporal properties of few TBHCs during their X-ray outbursts with TCAF will be discussed.

\end{abstract}


\section{Introduction}

Recently after the inclusion of TCAF model (Chakrabarti \& Titarchuk, 1995, hereafter CT95) into HeaSARC's 
spectral analysis software package XSPEC as an additive table model (Debnath et al., 2014, 2015a,b; 
Mondal et al. 2014; Jana et al., 2016; hereafter DCM14, DMC15, DMCM15, DMC14, JDCMM15 respectively), 
one can directly extract physical flow parameters, such as two types of accretion (Keplerian disk $\dot{m_d}$ 
and sub-Keplerian halo $\dot{m_h}$) rates and shock parameters (location $X_s$ and compression ratio $R$ of shock) 
from spectral fits. Evolution of these flow parameters, {\it accretion rate ratio} (ARR; i.e., $\dot{m_h}$/$\dot{m_d}$), 
and nature of quasi-periodic oscillations (QPOs; if present) provides us a better understanding on 
accretion flow dynamics around BHs.


\section{Results and Discussions}

Until now, we successfully studied accretion flow dynamics of four Galactic TBHCs (H~1743-322, GX~339-4, MAXI~J1659-152, 
MAXI~J1836-194) using RXTE data during their X-ray outbursts with TCAF.
A strong correlation between spectral and temporal properties of these TBHCs is observed. In ARR-intensity diagram 
(ARRID; see, MDC14, JDCMM15) different spectral states are found to be correlated with different branches of the diagram. 
Generally hard (HS), hard-intermediate (HIMS), soft-intermediate (SIMS) and soft (SS) spectral states 
are observed during an outburst of TBHCs in the sequence of HS $\rightarrow$ HIMS$ \rightarrow$ SIMS $\rightarrow$ SS 
$\rightarrow$ SIMS $\rightarrow$ HIMS $\rightarrow$ HS.
Our recent studies with TCAF confirms that the boundaries and properties of these different spectral states can easily 
be explained with the variation of ARR and nature of QPOs (if present). Also one can predict the frequency of the 
dominating QPOs with TCAF model fitted shock parameters (see, DCM14). We believe that an outburst is triggered due 
to a sudden rise in viscosity ($\alpha$ $\geq 0.15$, see for more details, Nagarkoti \& Chakrabarti, this Volume) 
and is turned off due to the reduction in viscosity. 
If viscosity does not rise significantly, Keplerian component does not become dominating over sub-Keplerian component, 
and as a result of that we may miss SS as well as SIMS during an outburst (see, DMCM15, JDCMM15), which could be 
termed as `failed' outburst. The nature of QPOs (generally type C) are found to evolve monotonically during 
rising/declining phase of HS and HIMS (Chakrabarti et al., 2008, 2015; Debnath et al., 2010, 2013; Nandi et al., 2012), with 
a local maxima of ARR on the transition day of HS $\rightleftharpoons$ HIMS and of evolving QPO frequency on 
HIMS$\rightleftharpoons$SIMS transition day (see, MDC14, DMC15, DMCM15). QPOs (generally type B or A) are observed 
sporadically on and off during SIMS and with absentia in SS (Nandi et al., 2012; Debnath et al., 2008, 2013). Generally, 
ARR rapidly increases during HS (rising), and decreases during HIMS (rising). The opposite scenario is observed during 
declining phase of these two states. It also has been found that ARR stays at lower values during 
SIMS (slow random variation) and SS (roughly constant).

\end{document}